\documentclass[10pt,conference]{IEEEtran}
\usepackage{cite}
\usepackage{amsmath,amssymb,amsfonts}
\usepackage{algorithmic}
\usepackage{graphicx}
\usepackage{textcomp}
\usepackage{xcolor,colortbl}
\usepackage{soul}
\usepackage{comment}
\usepackage[linesnumbered,ruled,vlined]{algorithm2e}
\usepackage{balance}
\usepackage{microtype}
\usepackage{multirow}
\usepackage[normalem]{ulem}
\useunder{\uline}{\ul}{}
\usepackage{booktabs}
\usepackage{setspace}

\usepackage{tikz}

\newcommand*\circled[1]{\tikz[baseline=(char.base)]{\small{\textbf{
			\node[shape=circle,fill,inner sep=0.75pt] (char) {\textcolor{white}{#1}};}}}}

\newcommand{\tool}{{\sc{HipHarness}}\xspace}

\definecolor{amethyst}{rgb}{0.6, 0.4, 0.8}
\newcommand{\yixue}[1]{{\color{cyan}(Yixue: #1)}}
\newcommand{\neno}[1]{{\color{red}(Neno: #1)}}

\newcommand{\marcelo}[1]{{\color{blue}(Marcelo: #1)}}

\newcommand{\cut}[1]{}
\def\BibTeX{{\rm B\kern-.05em{\sc i\kern-.025em b}\kern-.08em
    T\kern-.1667em\lower.7ex\hbox{E}\kern-.125emX}}
\begin{document}

\title{Assessing the Feasibility of Web-Request 
\\Prediction Models on Mobile Platforms}

\author{\IEEEauthorblockN{Yixue Zhao}
\IEEEauthorblockA{
University of Massachusetts Amherst\\
Amherst, USA \\
yixuezhao@cs.umass.edu}
\and
\IEEEauthorblockN{Siwei Yin}
\IEEEauthorblockA{
Beijing University of Posts and \\
Telecommunications, Beijing, China \\
ysw@bupt.edu.cn}
\and
\IEEEauthorblockN{Adriana Sejfia}
\IEEEauthorblockA{
University of Southern California\\
Los Angeles, USA \\
sejfia@usc.edu}
\and
\IEEEauthorblockN{Marcelo Schmitt Laser}
\IEEEauthorblockA{
University of Southern California\\
Los Angeles, USA \\
marcelo.laser@gmail.com}
\and
\IEEEauthorblockN{Haoyu Wang}
\IEEEauthorblockA{
Beijing University of Posts and \\
Telecommunications, Beijing, China \\
haoyuwang@bupt.edu.cn}
\and
\IEEEauthorblockN{Nenad Medvidovic}
\IEEEauthorblockA{
University of Southern California\\
Los Angeles, USA \\
neno@usc.edu}
}

\maketitle

\begin{abstract}
\looseness-1
Prefetching web pages is a well-studied solution to reduce network latency by predicting users' future actions based on their past behaviors.
However, such techniques are largely unexplored on mobile platforms. 
Today's privacy regulations 
make it infeasible to explore  prefetching with the usual strategy of amassing large amounts of data over long periods and constructing conventional, ``large'' prediction models. 
Our work is based on the observation that this may not  be necessary: Given previously reported mobile-device usage trends (e.g., repetitive behaviors in brief bursts), we hypothesized 
 that prefetching should work effectively with ``small'' models trained on mobile-user requests collected during much shorter time periods. 
 To test this hypothesis, we constructed a framework for automatically assessing  prediction models, and used it to conduct an extensive empirical study based on over 15 million HTTP requests collected from nearly 11,500 mobile users during a 24-hour period, resulting in over 7~million  models. 
Our results demonstrate the feasibility of prefetching with small models on mobile platforms, directly motivating future work in this area.
We further introduce several strategies for improving prediction models while reducing the model size.
Finally, our framework provides the foundation for future explorations of effective prediction models across a range of usage scenarios.
\end{abstract}


\section{introduction}
\label{sec:intro}

\looseness-1
Prefetching network requests is a well-established area aimed at reducing user-perceived latency~\cite{de2010referrer,gunduz2003web,ruamviboonsuk2017vroom,mickens2010crom,wang2016speeding,choi2018appx,zhao2018leveraging,wang2012far,palpanas1998webppm,padmanabhan1996usingDG,ban2007online}.
This work can be classified into two categories based on how predictions are made~\cite{ali2011survey}:
\emph{content-based} techniques predict future network requests by analyzing application content, such as web page structure, to anticipate ``sub-resources'' (e.g., images, JavaScript files) that will be needed by a web page; 
\emph{history-based} techniques predict future  requests by analyzing past requests.

Recent research has highlighted the opportunity to apply prefetching on mobile platforms~\cite{zhao2018empirically}. 
However, existing work has primarily explored {content-based} strategies that carry over several limitations~\cite{zhao2018leveraging,choi2018appx,malavolta2019nappa}.
First, their accuracy is impacted by the complexity of the program analyses  they employ. 
For instance, PALOMA \cite{zhao2018leveraging} and APPx \cite{choi2018appx} rely on static  analysis 
that is known to be unsound~\cite{wang2016unsoundness} and is thus guaranteed to miss certain prefetchable requests. 
Second, they rely on apps' logic and thus may not be effective for certain apps. For example, PALOMA analyzes the app to determine a request's URL one callback before when the request is issued, and then prefetches it; if URLs are unknown one callback ahead, PALOMA will never prefetch.
Finally, content-based techniques do not generalize across different frameworks (e.g., Android, iOS) or app types (e.g., native apps vs. mobile browsers).

\looseness-1
{By contrast, history-based} prefetching 
does not require program analysis and is platform-independent: 
\cut{unlike content-based prefetching, which is tied to a specific platform and needs constant adjustment as the app evolves, history-based prefetching}
it only relies on  past requests.  
However, despite the large body of work targeting history-based prefetching in browsers~\cite{de2007web,de2010referrer,davison2002predicting,gunduz2003web,palpanas1998webppm,padmanabhan1996usingDG,ibrahim2000neural,pitkow1999mininglongestrepeatin,ban2007online,yang2004building}, such techniques remain largely unexplored on mobile platforms.

\looseness-1
This gap is caused by a key challenge in applying history-based prefetching today: {the user data is hard to obtain} 
due to the profusion of data-privacy regulations introduced and regularly tightened around the world~\cite{privacylaw,Leu2019Jan,consumersinternational,PricewaterhouseCoopers2020Apr,DataProtectionLaw}. 
Although such regulations vary across regions, the global trend is to restrict the data collection to only the necessary amount with user consent (e.g., see GDPR \cite{consumersinternational}).
Furthermore, today's users are increasingly protective of their data and much more likely to agree to its collection if it is restricted to small amounts \cite{Ovide2020Jul,Swant2019Aug,BibEntry2018Mar}.
This makes the traditional strategy applied
 in this area, which has relied on “large” data impractical today.
For example, a study published in 2011 reported on data collected from 25 iPhone users over the course of an entire year~\cite{shepard2011livelab} and inspired a number of follow-up studies~\cite{wang2012far,yan2012fast,mohan2013prefetching,lymberopoulos2012pocketweb,ali2011survey,likamwa2011can,rahmati2012exploring}. 
A decade later, such~protracted data collection would be difficult to imagine, both because it would likely fall afoul of legal regulations introduced in the meantime and because today's end-users are  more keenly aware and protective of their data. 

Given these constraints, an obvious strategy would be to limit the amount of  data on which the prefetching relies. 
However, there is no evidence that it is  feasible to predict future requests based on \emph{small amounts of data}.
This appears to have been an important factor that has discouraged the exploration of history-based prefetching in recent years.
We believe that this has been a missed opportunity. 
Namely, the previously reported  mobile-device usage patterns---repetitive activities in brief bursts~\cite{guo2017looxy,shoukry2013evolutionary,D.W.2018Determining,Sarker2018Individualized,Bulut2015Understanding}---lead us to hypothesize that history-based prefetching may work effectively with \emph{small prediction  models} trained on mobile-user requests collected during \emph{short time periods}. 
If borne out in practice, this would open new research avenues. 

To evaluate our hypothesis, and to facilitate future explorations in this area, we first construct a tailorable framework \tool, which provides several customizable components to \emph{automatically assess prediction models}  across a range of scenarios. 
For example, \tool  allows comparing the effectiveness of different prediction algorithms by running them side-by-side on the same data, measuring the impact of different training-data sizes on accuracy, and so on. 

\looseness-1
\tool enables us to flexibly  assess prediction models built with any algorithm of interest.
In this paper, we specifically customize \tool to analyze models built with the three most widely employed history-based prediction algorithms from the traditional browser domain~\cite{padmanabhan1996usingDG,palpanas1998webppm,Bouras2004MP}, as well as a fourth algorithm we  introduce to serve as the evaluation baseline. Our study uses the mobile-network traffic obtained from nearly 11,500 users at a large university during a  {24-hour period}. 
 The selection of this time period was guided by  previously made observations of repetitive mobile-user behaviors during a single day~\cite{shoukry2013evolutionary,D.W.2018Determining,Sarker2018Individualized,Bulut2015Understanding}. 
The closest study to ours~\cite{wang2012far} only evaluates one algorithm~\cite{Bouras2004MP} with mobile-browser data collected over a year from 25 iPhone-using undergraduates at Rice university~\cite{shepard2011livelab}. 
By comparison, our study evaluates four  algorithms on both mobile-browser and mobile-app data, relying on $\approx$400$\times$ more users from a much more diverse user base, but during a $\approx$400$\times$ shorter time frame.

Our dataset comprises over 15 million HTTP requests from nearly 31,000 Internet domains, allowing us to explore orders-of-magnitude more models compared to prior work. 
We use \tool to assess over 7 million models tailored to each user based on their past usage, varying from a single request to over 200,000 requests. 
While \tool allows the exploration of various research questions, this paper targets
\circled{1}~\emph{the repetitiveness of user requests during short time periods} as the key prerequisite for the feasibility of small prediction models; \circled{2}~\emph{the effectiveness of existing prediction algorithms on mobile platforms} to provide insights for future algorithms; and \circled{3}~\emph{strategies for reducing training-data sizes without sacrificing model accuracy} as a way of yielding even smaller models. 

This paper makes the following contributions.

	\begin{enumerate}
	\item We design and implement \tool, a customizable framework for automatically exploring history-based prefetching across a wide range of scenarios.
	\item We conduct the first extensive study of history-based prefetching on mobile platforms, providing empirical evidence on the feasibility of small prediction models, and in turn, opening up a new research direction.
	\item We demonstrate that existing prediction algorithms are promising starting points for developing prefetching solutions, and provide insights on their further optimization.
	\item We develop concrete strategies for bounding the training-data sizes without sacrificing the models' accuracy. 
	\item We provide our artifacts to foster  future studies in this area, including \tool's source code, the evaluation results of  over 7 million prediction models we built, and the scripts for analyzing the results~\cite{repo}.
\cut{We provide the implementation of \tool, including several reference components, directly  fostering future studies in this area.}
	\end{enumerate}

Section~\ref{sec:foundation} describes the related work and the foundation of our study. Section~\ref{sec:appr} introduces our  framework, \tool. 
 Section \ref{sec:study} overviews our empirical study enabled by \tool. 
 Section \ref{sec:results} discusses our findings as well as the study's validity threats. 
 Section~\ref{sec:conclusion} concludes the paper. 

\section{Related Work and Foundation}
\label{sec:foundation}

\looseness-1
This section discusses the related work and background of {history-based} prefetching, followed by the rationale behind the selection and evaluation of the algorithms used in our study. 


\subsection{Related Work}
\label{foundation:related}

Prefetching has yielded a large body of work since the birth of the Internet~\cite{de2007web,de2010referrer,davison2002predicting,gunduz2003web,ruamviboonsuk2017vroom,mickens2010crom,wang2016speeding,choi2018appx,zhao2018leveraging,wang2012far,palpanas1998webppm,padmanabhan1996usingDG,ibrahim2000neural,pitkow1999mininglongestrepeatin,ban2007online,yang2004building}, and has been shown promising on mobile platforms recently~\cite{zhao2018empirically}.
Related work includes identifying performance bottlenecks~\cite{nejati2016depth,wang2013demystifying,vesuna2016caching}, balancing trade-offs between prefetching benefits and waste~\cite{higgins2012informed,baumann2017every}, and {content-based} techniques  ~\cite{zhao2018leveraging,choi2018appx,malavolta2019nappa,wang2016speeding} as discussed earlier. 

\looseness-1
History-based prefetching has remained largely unexplored on mobile platforms.
Prior work has mainly focused on mobile browsers \cite{Kovvali2011Jul,weber2010mobile,Jin2007,Branch2011Nov,brown2016pre,lymberopoulos2012pocketweb}, while missing mobile apps where users spend over 80\% of their time~\cite{appdominant}.
The few techniques targeting mobile apps are restricted to specific domains, such as social media~\cite{wang2015earlybird}, video streaming~\cite{koch2014optimizing,siris2013improving}, and ads~\cite{mohan2013prefetching}.
We believe this restriction is caused by the already mentioned challenge of accessing user data, limiting history-based prefetching to specific applications and/or domains with public data (e.g., Twitter~\cite{wang2015earlybird}). 
To mitigate this challenge, our work explores a novel strategy of assessing history-based prefetching across different domains, by relying on \emph{small amounts of user data}.
The closest  work to ours is the already discussed evaluation of one history-based prediction algorithm (MP~\cite{Bouras2004MP}) using data collected during one year from 25 iPhone users~\cite{wang2012far}. 
By contrast, our conclusions are based on four algorithms, including MP, and a much larger user base but over a much shorter time period. 

\subsection{Prefetching Workflow}
\label{foundation:phases}
History-based prefetching consists of three  phases: \emph{training} the prediction model based on the historical data; \emph{predicting} future requests based on the trained model; and \emph{prefetching} requests based on the prediction results and runtime conditions. 


 In the \textbf{\textit{training}} phase, a prediction model is trained with past requests 
 to capture their relationship based on a  prediction algorithm (e.g., the probability of making a given request next).
One prediction algorithm can be used to produce multiple prediction models by taking as input different training data, such as different numbers of past requests.
In the \textbf{\textit{predicting}}  phase, the trained  model is used to predict future requests based on the current context (e.g., the current   request) 
and the prediction algorithm. 
Finally, in the \textbf{\textit{prefetching}} phase, certain predicted requests are prefetched  based on  runtime conditions, such as battery life and cellular-data usage  \cite{higgins2012informed}. 

\subsection{Prediction Algorithms}
\label{foundation:algorithm}

Our study's goal is to assess the applicability of existing prediction algorithms that are most likely candidates for effective use on mobile platforms.
The algorithm selection is guided by our focus on small  models. 
We thus exclude algorithms that rely on large amounts of training data, such as those using neural networks (e.g., LSTM \cite{sak2014LSTM}).
We specifically include the MP algorithm \cite{Bouras2004MP} since it is used in the closest related study \cite{wang2012far}.
Finally, we select two additional algorithms---DG \cite{padmanabhan1996usingDG} and PPM \cite{palpanas1998webppm}---that are widely used to predict web requests~\cite{domenech2006metric,domenech2010using,ossa2007improving,liu2016request,nanopoulos2003data,gellert2016web,cui2020improved}.
Note that there are multiple variations of the three selected algorithms~\cite{domenech2006metric,domenech2010using,domenech2006ddg,ossa2007improving,liu2016request,nanopoulos2003data,chen2002popularity,liu2009web,ban2007online}; a detailed summary can be found in a survey~\cite{ali2011survey}.
The variations in question target specific characteristics of traditional browsers, such as web-page structure \cite{domenech2006ddg}, and do not carry over to mobile platforms.
We thus rely on the originally defined  algorithms.
\cut{, and to motivate future work on improving them based on the characteristics of  mobile platforms.}


\textbf{Most-Popular} (MP) \cite{Bouras2004MP} maintains a list of the most-popular subsequent requests for each request in the training set. In the training phase, MP adds next requests within a specified window to the current request's list and stores their occurrences. In the predicting phase, MP predicts the most-popular requests. 

\textbf{Dependency Graph} (DG) \cite{padmanabhan1996usingDG} 
trains a directed graph indicating the dependencies among the  requests, where nodes  represent requests and arcs  indicate that a target node is visited after the original node within a window. Each arc has a weight that represents the probability that the target node will be visited next. 
In the predicting phase, DG predicts future requests based on their probabilities stored in the dependency graph.

\looseness-1
\textbf{Prediction-by-Partial-Match} (PPM)  \cite{palpanas1998webppm} uses a high-order Markov model that is context-sensitive since it considers the order of  requests. In the training phase, it builds a trie structure~\cite{trie} that indicates the immediate-followed-by relationship of the past requests. In the predicting phase, it predicts future requests whose parents match the previous $N$ requests.

\subsection{Study Focus}
\label{foundation:focus}

We focus on the \emph{accuracy} of prediction models built in the first two phases of the  workflow, i.e., \emph{training} and \emph{predicting}.
The third, \emph{prefetching} phase involves trading off runtime conditions, studied by a complementary body of research~\cite{higgins2012informed,baumann2017every}.
We exclude such runtime factors since they would taint the results of the models' accuracy (e.g., failing to prefetch requests due to a third-party server's errors).
Moreover, the expiration of the prefetched requests may vary depending on when the experiments are conducted, which would introduce additional bias into our results. 
Currently, determining whether a request has expired remains an open challenge as the HTTP headers~\cite{httpHeader}   (e.g., {Cache-Control}, {Expires}) are not trustworthy \cite{zhao2018empirically}. 
Thus, to eliminate runtime variations and fairly assess the models' accuracy, we assume ideal runtime conditions: each predicted request will be prefetched and will not expire.

As mentioned earlier, our prediction models are tailored to \emph{individual users} based on their past behaviors.
This is motivated by today's stringent regulations limiting personal data access.
This has shifted recent research to ``on-device'' prediction, to avoid exporting sensitive user data to servers~\cite{hard2018federated,lim2019federated,ramaswamy2019federated}.
At the same time, this trend of client-side prediction highlights the need for \emph{small prediction models} since mobile devices are resource-constrained.
We thus also assess the \emph{resource consumption} of different prediction models trained on-device. 

\section{The \tool Framework}
\label{sec:appr}

This section describes the design of \tool, a tailorable framework for automatically assessing  history-based prediction models, followed by the details of its instantiation. 

\subsection{\tool's Design}
\label{bg:framework} 

\begin{figure}[t]
  \centering
  \includegraphics[width=0.475\textwidth]{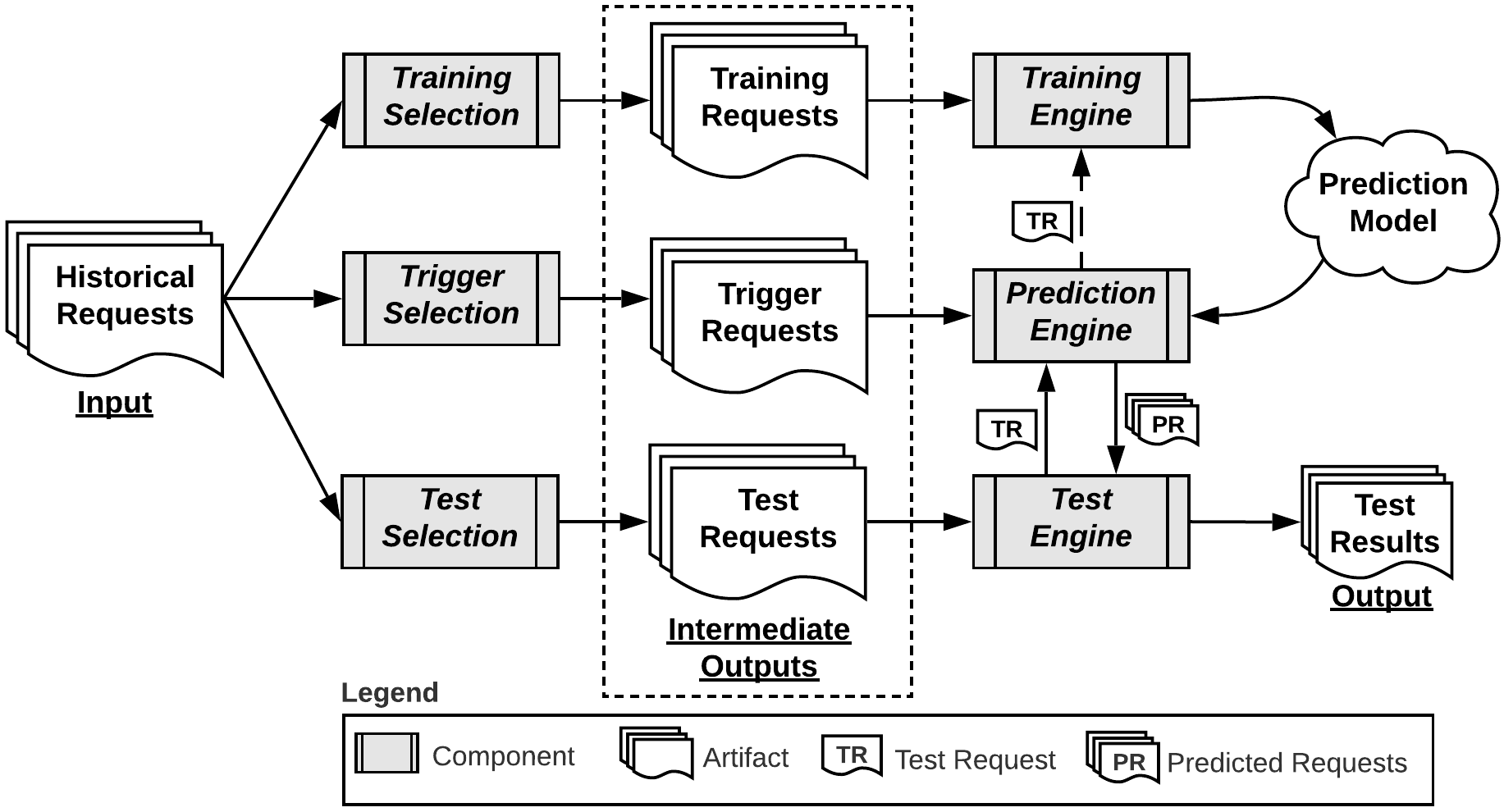}
  \vspace{-3mm}
  \caption{\tool's workflow for assessing history-based prediction models}
  \vspace{-4mm}
  \label{fig:testing_workflow}
\end{figure}

Figure \ref{fig:testing_workflow} depicts \tool's workflow, comprising six customizable  components that can be  reused, extended, or replaced as needed.
For instance, one can explore how much training data to use with a given prediction algorithm by varying  \textit{Training Selection} and reusing the remaining components.


\tool takes {\small \textsf{\small \textsf{Historical Requests}}} as input. Each component  (shaded boxes) are either reused from existing components, or provided anew if unavailable.
{\small \textsf{Historical Requests}} are past requests used to generate the intermediate outputs of {\small \textsf{Training Requests}}, {\small \textsf{Trigger Requests}}, and {\small \textsf{Test Requests}} based on the logic defined in \textit{Training Selection}, \textit{Trigger Selection}, and \textit{Test Selection}, respectively. 
{\small \textsf{Training Requests}} are used to build the {\small \textsf{Prediction Model}}. 
{\small \textsf{Trigger Requests}} triggers the prediction of subsequent requests, such as the current request. 
Finally, {\small \textsf{Test Requests}} are the future requests used to evaluate the {\small \textsf{Prediction Model}}; they represent the ``ground truth''. 

\looseness-1
The three intermediate outputs make the prediction and produce the final {\small \textsf{Test Results}} to evaluate the {\small \textsf{Prediction Model}}. 
The \textit{Training Engine} implements the algorithm used to train a specific {\small \textsf{Prediction Model}}, such as the dependency graph in DG~\cite{padmanabhan1996usingDG}, based on the selected {\small \textsf{Training Requests}}. 
Optionally, the trained {\small \textsf{Prediction Model}} can be updated dynamically (dashed arrow) while being evaluated: 
{\small \textsf{Test Requests}} become historical requests after being tested, and can be used to train the {\small \textsf{Prediction Model}}. 
 \textit{Prediction Engine}  implements the algorithm that predicts subsequent requests based on {\small \textsf{Trigger Requests}} and the trained {\small \textsf{Prediction Model}}. 
Finally, when evaluating the {\small \textsf{Prediction Model}},  \textit{Test Engine} iteratively invokes  \textit{Prediction Engine} with a series of  {\small \textsf{Test Requests}}.  \textit{Prediction Engine} selects  $N$ requests that immediately precede each {\small \textsf{Test Request}}  to obtain a set of {\small \textsf{Predicted Requests}}. 
 \textit{Test Engine} then compares the {\small \textsf{Predicted Requests}} with the corresponding {\small \textsf{Test Requests}} (i.e., ``ground truth'') 
and outputs the {\small \textsf{Test Results}} that contain the  information needed to calculate the evaluation metrics of interest (see Section~\ref{sec:study:metric}). 

\looseness-1
\tool supports ``plug and play'' 
by tuning and/or replacing each of the   components to explore different research questions. 
The \emph{Training Selection}, \emph{Trigger Selection}, and \emph{Test Selection} components can be customized based on different selection strategies. For example, different parts of  {\small \textsf{Historical Requests}} can be chosen based on  desired ratios or time periods, to produce the corresponding {\small \textsf{Training Requests}}, {\small \textsf{Trigger Requests}}, and {\small \textsf{Test Requests}}. 
Likewise,  \emph{Training Engine} and \emph{Prediction Engine} can be tailored with specific algorithms  to train the  models based on the selected {\small \textsf{Training Requests}}. 
Finally,  \emph{Test Engine} can be customized to evaluate different prediction models with various testing strategies. 
For instance, as discussed earlier, \emph{Test Engine} may enable updating the prediction model dynamically; 
by plugging-in \emph{Test Engine}s that implement different dynamic-update strategies, \tool can isolate the impact of those strategies by producing their  side-by-side {\small \textsf{Test Results}}. 
\subsection{\tool's Instantiation}
\label{data:implementation}

We have instantiated \tool by implementing several variations of its six components. 
Instances of \textit{Training Selection} and \textit{Test Selection} are implemented to select requests based on different ratios, such as using the first 80\% of the {\small \textsf{Historical Requests}} as {\small \textsf{Training Requests}} and the remaining 20\% as {\small \textsf{Test Requests}}.
\textit{Trigger Selection} is implemented to select one current request to trigger the prediction of subsequent requests.
Four pairs of \textit{Training Engine} and \textit{Prediction Engine} are implemented based on the three algorithms introduced in Section \ref{sec:foundation}---DG \cite{padmanabhan1996usingDG}, PPM \cite{palpanas1998webppm}, and MP~\cite{Bouras2004MP}---and a fourth baseline algorithm we will describe in Section \ref{sec:results}.
\cut{Initially, the {\small \textsf{Prediction Model}} is trained by  the \textit{Training Engine} based on the {\small \textsf{Training Requests}} and is fed to  the \textit{Prediction Engine}. While evaluating the  results of the {\small \textsf{Prediction Model}}, \textit{Test Engine} initiates  dynamic updates of the model by invoking  \textit{Prediction Engine}, as shown in Algorithm~\ref{alg:test_engine}.}
Finally, \emph{Test Engine}'s implementation is detailed in Algorithm \ref{alg:test_engine}.

\begin{algorithm}[b]
\small
\begin{spacing}{0.9}
\DontPrintSemicolon 
\KwIn{PredictionEngine $PE$, TestRequests $test\_reqs$}
\KwOut{$cache.$\textsc{Size}, $hit\_set$, $miss\_set$, \#$prefetch$, \#$hit$, \#$miss$}

$cache = \emptyset$, $hit\_set = \emptyset$, $miss\_set = \emptyset$, \\
\#$prefetch = 0$, \#$hit = 0$, \#$miss = 0$

\ForEach{$current\_req \in  test\_reqs$}{
		$predicted\_reqs \gets PE.\textsc{Predict}(current\_req.pre)$\\
		\ForEach{$candidate \in predicted\_reqs$}{
			\If{$\neg$\textsc{IsCached}($candidate$)}{
				\textsc{Prefetch}($candidate$)\\
				$cache.$\textsc{Put}($candidate$)\\
				$prefetch \gets prefetch + 1$
			}
		}
		\If{\textsc{IsCached}($current\_req$)}{
			$hit \gets hit + 1$ \\
			$hit\_set.$\textsc{Put}($current\_req$)
		}
		\Else{
			$miss \gets miss + 1$ \\
			$miss\_set.$\textsc{Put}($current\_req$)
		}
		$PE.$\textsc{update\_Model}($current\_req$)
	}
\end{spacing}
\vspace{-1mm}
\caption{\sc Test Engine}
\label{alg:test_engine}
\end{algorithm}

\textit{Test Engine}'s objective is to output  information needed for evaluating \textit{Prediction~Engine}'s (\textit{PE}) results. 
For each request $current\_req$ in {\small \textsf{Test Requests}} $test\_reqs$, \textit{PE} predicts the potential current requests based on $current\_req$'s previous request $current\_req.pre$ (Line 4)
and prefetches the predicted requests that are not already in the cache (Lines 5-9).

\looseness-1
As discussed in Section~\ref{foundation:focus}, we 
assume that all  predicted requests should be prefetched and will not expire. 
We thus place the predicted requests in the cache without sending them to the server 
to avoid tainted results caused by spurious runtime variations.
The cache size is unbounded in our study:  since we focus on small amounts of historical data guided by our hypothesis (recall Section~\ref{sec:intro}), the required cache size is negligible compared to the available storage on mobile devices.

 \textit{Test Engine} then compares ${current\_req}$ with the cached  requests
to determine whether it can be reused, and updates the corresponding information (Lines 10-15).
Finally, $PE$ dynamically updates the {\small \textsf{Prediction Model}} by adding  ${current\_req}$ to train the model (Line~16) since ${current\_req}$  becomes a past request as \textit{Test Engine} moves to the next request in  ${test\_reqs}$.

In the end, \textit{Test Engine} outputs the size of the {cache} ($cache.$\textsc{Size});
 unique prefetched requests that were subsequently used (${hit\_set}$);
 unique requests in {\small \textsf{Test Requests}} that were not in the cache when requested (${miss\_set}$);
 total number of prefetched requests (${\#prefetch}$); and
 numbers of times a request was in the cache (${\#hit}$) vs. not in the cache (${\#miss}$) when requested. 
We use this  information to define metrics that evaluate the accuracy of prediction models in Section~\ref{sec:study}.

\section{Empirical Study Overview}
\label{sec:study}

This section provides the details of our empirical study enabled by \tool, including the research questions we focus on, the data collected, and the evaluation metrics used.

\subsection{Research Questions}
\label{sec:study:rq}

\cut{As discussed in Section~\ref{sec:intro}, our study is guided by the challenge of obtaining user data,
which motivated us to investigate {small prediction models} built with data  collected during {short time periods}. 
Given previously made observations of repetitive mobile-user behaviors in brief bursts~\cite{shoukry2013evolutionary,D.W.2018Determining,Sarker2018Individualized,Bulut2015Understanding},}

\looseness-1
An overarching hypothesis frames our study: History-based prefetching may work effectively with \emph{small prediction  models} trained on mobile-user requests collected during \emph{short~time periods}. 
To evaluate it, we focus on three research questions.

\begin{itemize}
	\item \textbf{RQ$_1$} -- To what extent are mobile users' requests repetitive during short time periods?
	\item \textbf{RQ$_2$} -- How effective are the existing prediction algorithms when applied on mobile platforms using small prediction models?
	\item \textbf{RQ$_3$} -- Can the training-data size be reduced without significantly sacrificing the prediction models' accuracy?
\end{itemize}

\subsubsection*{\textbf{RQ1 -- Repetitiveness of  User Requests}}
\label{sec:rq:rq1}

\looseness-1
Since history-based techniques can only predict requests that have appeared in the past, our study is centered on \emph{repeated requests}~\cite{lymberopoulos2012pocketweb}, i.e., identical requests issued by a user.
Requests are considered identical when they have the same URLs, including \texttt{GET}  parameters~\cite{rfc2616get}.
We assume the requests will not expire by design as discussed in Section \ref{foundation:focus}.
Specifically, we explore the extent to which mobile users send repeated requests during a  day, aiming to understand the ``ceiling'' that can be achieved by any history-based techniques with small models.
Note that prior work has only reported the repetitiveness of coarse-grained  behaviors  (e.g., phone-calls~\cite{Sarker2018Individualized}, mobility patterns \cite{tseng2006efficient,hsu2007modeling,yang2015characterizing}). 

\subsubsection*{\textbf{RQ2 -- Effectiveness of Prediction Algorithms}}
\label{sec:rq:rq2}

\looseness-1
In the closest study, Wang et al.~\cite{wang2012far} found history-based algorithms ineffective.
However, their study was conducted a decade ago, had a limited scope, and relied on the conventional, \emph{large} prediction models. 
We re-assess their conclusions using  
small models, while extending the study's scope in three dimensions.
First, Wang et al. analyzed 25 iPhone users who were undergraduates; we rely on 11,476 users with different types of mobile devices and diverse occupations. 
Second, the their dataset only included requests collected from the Safari browser; 
 we rely on both mobile-browser and app data to more comprehensively cover mobile-user behaviors. 
Finally, only one history-based algorithm  (MP~\cite{Bouras2004MP}) was previously investigated; our conclusions are based on the three most widely-employed algorithms, including MP, and an additional baseline algorithm. 

\noindent\subsubsection*{\textbf{RQ3 -- Reducing Training-Data Size}}
\label{sec:rq:rq3}
We explore whether the amount of training data can be reduced without sacrificing accuracy.
Guided by Pareto principle~\cite{8020}, we posit that top-20\% of the training data  are likely to account for 80\% of the results, 
and explore three strategies to select the training data:
\circled{1}~\emph{Most Occurring Requests --} 
Given a request sequence in the training set, we group repeated requests, and 
use the requests in the largest 20\% of the groups to train the models. 
\circled{2}~\emph{Most Accessed Domains --} We  group  requests that belong to the same domain  (e.g., {google.com/*}), and  train the models with the requests in the largest 20\% of the groups. 
\circled{3}~\emph{Most Suitable Domains --} Domains that tend to contain repeated requests are potentially good candidates for prefetching. 
Thus, we again group the requests in the same domain together. This time, we rank the groups by the \emph{proportion} of  repeated requests, and use the top-20\% of the groups to train the prediction models.

We further investigate whether there is a \emph{lower-bound} on the training-data size that yields the smallest model without sacrificing the resulting  accuracy.
To that end, we designed the \emph{Sliding-Window} approach (see Section~\ref{sec:results}) to study the impact of different numbers of requests in the training set on the prediction accuracy, and to  
identify the lower-bound. 

\subsection{Data Collection}
\label{sec:study:data}


\textbf{Data Collection Process.} The network traces were collected 
at the gateway between Beijing University of Posts and Telecommunications's campus network and the Internet. 
The measurement servers were 
placed at the campus gateway router, connected via optical splitters to the Gigabit access links.
The HTTP headers were captured by the servers along with the timestamps. The authentication information (User ID) identifies the traffic from the same user. 

\textbf{Ethical Considerations.} Our study was approved by the university, and was guided by the agreement the authors signed.
The study strictly followed the Research Data Management Policy of the university, including data storage, sharing, and disposal. 
All  collected raw data was processed by the university's network center. All recorded IP addresses and authentication information were anonymized. The authors did not have access to any of the raw data at any time. 
Unlike prior work that selected 25 users~\cite{wang2012far,shepard2011livelab}, our data includes all users who accessed the campus network without any selection. 

\looseness-1
\textbf{Dataset Overview.}
Our study  
aims to assess small models trained on mobile-user requests collected during  much shorter time periods compared to the conventional weekly or monthly models~\cite{wang2012far, davison2002predicting,gunduz2003web,tuah2000performance,de2010referrer,de2007web}.
To that end, we were given access to the network traffic collected by the university, spanning the 24 hours of May 28, 2018 (a randomly selected date). 
This included the traffic from nearly 11,500 accounts, representing all users who accessed the campus network via a mobile device\footnote{PCs and mobile devices use different clients for authentication. We only collected the mobile-network traffic based on the 
 authentication information.} during that time: students, faculty, staff, contract employees, residents, outside vendors, and visitors. These users exhibit various behavior patterns and form a diverse group for our study.  
We further filtered the mobile-network traffic to  include only the  requests involving the HTTP \texttt{GET}  method~\cite{rfc2616method}: \texttt{GET} requests are considered ``safe'' for prefetching in that they do not have any side-effects on the server~\cite{rfc2616get,rfc2616safe,zhao2018empirically}. Ultimately, we collected 15,143,757 \texttt{GET} requests from 11,476 users. 
Each request is identified by its URL, including \texttt{GET} parameters if any.

\vspace{-1mm}
\subsection{Evaluation Metrics}
\label{sec:study:metric}

As discussed in Section~\ref{foundation:focus}, we focus on the \emph{accuracy} of the prediction models.
We thus leverage two widely adopted accuracy metrics in the prefetching literature, and introduce a new  accuracy metric.
The metrics are computed based on the information output by our \textit{Test Engine} (recall Algorithm \ref{alg:test_engine}). 

\textbf{Static Precision} measures the percentage of  correctly predicted \emph{unique} requests, i.e., the ratio of prefetched requests that are subsequently used to the number of all prefetched requests \cite{domenech2006metric}. This metric is often referred to as \emph{precision} or \emph{hit ratio} in the literature~\cite{Bouras2004MP,ibrahim2000neural,Markatos1996AT1,wang2012far,palpanas1998webppm,nanopoulos2001effective, domenech2006metric}. 
\begin{equation*}
\vspace{-1mm}
    Static\ Precision = \frac{|hit\_set|}{\#prefetch}
\end{equation*}

\textbf{Static Recall} is a new metric we introduce as the counterpart to Static Precision. Static Recall measures the ratio of  \emph{unique} requests that were previously prefetched (and have been cached) to the total number of unique requests made by a user. 
\begin{equation*}
\vspace{-1mm}
    Static\ Recall = \frac{|hit\_set|}{|hit\_set| + |miss\_set|}
\end{equation*}


\textbf{Dynamic Recall} measures the ratio of previously-prefetched requests  to all requests a user issues~\cite{domenech2006metric}. This metric is often referred to as \emph{recall} or \emph{usefulness} in the prefetching literature~\cite{Bouras2004MP,ibrahim2000neural,Markatos1996AT1,wang2012far,palpanas1998webppm,nanopoulos2001effective, domenech2006metric}. 
\begin{equation*}
\vspace{-1mm}
    Dynamic\ Recall = \frac{\#hit}{\#hit + \#miss}
\end{equation*}

We do not use a dynamic counterpart to Static Precision 
because it is not meaningful.
To measure the ratio of correctly predicted requests dynamically, this metric needs to reward the predicted requests ({hits}) \emph{each time} they are accessed, 
and penalize ``useless'' requests that were prefetched but  never used. 
However, the reward for {hits} would  be potentially  unbounded. 
Moreover, the metric would allow cases in which a model with a low Static~Precision has an artificially high Dynamic Precision: 
repeated {hits} of the same request would progressively diminish the  penalty for arbitrarily many useless requests. 

\section{Results and Lessons Learned}
\label{sec:results}


This section presents the results of our  empirical study performed using \tool, and its takeaways. 

\looseness-1
As mentioned earlier, our study is based on over 15~million HTTP \texttt{GET} requests, reflecting the mobile traffic collected from 11,476 users at a large university. 
Each user  sent 1,320 requests on average during the {single day}. This is over 100$\times$ more than what was reported a decade ago, where each user sent 4,665 requests for the  \emph{entire year}~{\cite{wang2012far,shepard2011livelab}}, 
reinforcing the impracticality of traditional large models for mobile platforms. For instance, the  strategy  of building monthly models~\cite{wang2012far} would encompass $\approx$40,000 requests if applied on our dataset.
\looseness-1
Starting with this initial user set, we filtered out outlier users identified by box-and-whisker plot method~\cite{tukey77} based on the number of requests they send, since these users do not represent typical mobile-device usage. 
This process resulted in the final dataset of 9,900 subject users as shown in Table \ref{tbl:req_per_user}. 
\cut{Table \ref{tbl:req_per_user} shows the numbers of requests per user in our initial  and  final datasets.} 

\begin{table}[t]
\centering
\caption{Requests per user}
\label{tbl:req_per_user}
\vspace{-3mm}
\tiny
\centering
\resizebox{.8\linewidth}{!}{
\begin{tabular}{|c|c|c|c|c|c|}
\hline
\multicolumn{3}{|c|}{\textbf{Initial 11,476 Users}} & \multicolumn{3}{c|}{\textbf{Final 9,900  Users}} \\ \hline
Min        & Avg       & Max       & Min          & Avg         & Max        \\ \hline
1               & 1,320        & 235,837         & 10                & 981           & 3,923            \\ \hline
\end{tabular}
}
\vspace{-3mm}
\end{table}

\subsection{RQ1 -- User-Request Repetitiveness}
\label{sec:result:predictability}


\begin{itemize}
	\item  \textbf{RQ$_1$} -- To what extent are mobile users' requests repetitive during short time periods?
\end{itemize}


\begin{table}[b]
\vspace{-6mm}
\centering
\caption{Repeated requests across users}
\vspace{-3mm}
\label{tbl:repeated_percentage}
\centering
\tiny
\resizebox{0.9\linewidth}{!}{
\begin{tabular}{|c|c|c|c|c|c|c|c|}
\hline
\multicolumn{4}{|c|}{\textbf{Percentage}} & \multicolumn{4}{c|}{\textbf{Number}} \\ \hline
Min   & Avg    & Max    & SD     & Min  & Avg  & Max    & SD   \\ \hline
0\%   & 28\%   & 98\%   & 17\%   & 0    & 293  & 3,225  & 353  \\ \hline
\end{tabular}
}
\end{table}

To answer {RQ$_1$}, we calculate both the {number} and {percentage} of \emph{repeated requests} in each user's  data, 
as they indicate different aspects of predictability. The \emph{percentage} provides insights about the potential cost: low percentage  indicates high proportion of non-predictable requests, increasing the cost of building the model. The \emph{number} sets the upper-bound on  requests that can be prefetched. 

\looseness-1
Table~\ref{tbl:repeated_percentage} shows this data  
across the 9,900 subject users. 
The minimums indicate that certain users did not access the same URL more than once. 
 222 (2\%) of our  users fall in this category. 
Further investigation uncovered that these users do not  access the network frequently enough to show repetitive behaviors: the average number of the network requests sent by these users is  32, i.e., $\approx$1 request every 80 minutes during the 24 hours. 

\looseness-1
On average, 28\% of the requests are repeated. An average user sends 293 repeated requests, which can be prefetched and reused from a cache. 
The maximum percentage (98\%) and number (3,225) show that history-based prefetching   can especially benefit certain users. 
We also find a large variation across different users (see standard deviation in Table \ref{tbl:repeated_percentage}). 
This  indicates that individual users exhibit markedly different behaviors, which reinforces our choice of building personalized prediction models for each user on the client side (recall Section~\ref{foundation:focus}). 



To provide further insights for future techniques, we study the characteristics of \emph{frequently repeated requests} by 
computing how many times each repeated request occurs per user. 
Due to space constraints, we highlight two  findings. 
\circled{1} The average numbers of certain users' repeated requests are unusually high, with the maximum of 779. 
Further investigation showed that these users tended to send large numbers of requests to obtain the same WiFi configuration files from a specific domain. Since the data  available to us was sanitized, we can only 
hypothesize that this was due to server-side flaws. 
\circled{2} The maximum occurrence of a repeated request across all users was 2,634.
This and other high values corresponded
to continually obtaining information from certain servers based on the given users' unchanging location coordinates, as indicated in \texttt{GET} requests' parameters. 
Such domains may also have flaws that require resending the user location even when it remains the same.
Both instances clearly point to opportunities for caching. 

\subsection{RQ2 -- Prediction-Algorithm Effectiveness}
\label{result:rq2}

\begin{itemize}
	\item \textbf{RQ$_2$} -- How effective are the existing prediction algorithms when applied on mobile platforms using small prediction models?
\end{itemize}

RQ$_1$'s results indicate that our 24-hour dataset is amenable to prefetching, motivating us to 
 apply  the existing prediction algorithms on it. 
The rationale behind our algorithm selection was discussed in Section \ref{foundation:algorithm}.
We further develop an additional algorithm (\emph{Na\"ive}) to serve as the evaluation baseline. 
Na\"ive assumes that each request appeared in the past will appear again, and caches each such request.
Na\"ive thus guarantees the upper-bound on the number of predictable requests, and provides a baseline for measuring any other prediction algorithm's recall. 

To answer {RQ$_2$}, we use \tool to obtain the {\small \textsf{Test Results}} (recall Figure~\ref{fig:testing_workflow}) of the  prediction models built for each  user.
As discussed in Section \ref{data:implementation}, we implemented four pairs of \emph{Training Engine} and \emph{Prediction Engine}  in \tool, based on the three  existing algorithms discussed in Section~\ref{foundation:algorithm} and Na\"ive. 
In the end, \emph{Test Engine} outputs the {\small \textsf{Test Results}} needed for evaluating the models' accuracy (recall  Algorithm~\ref{alg:test_engine}), using the three accuracy metrics discussed in Section~\ref{sec:study:metric}.

For all four algorithms, we follow the common approach of selecting the first 80\% of {\small \textsf{Historical Requests}} as {\small \textsf{Training Requests}}, and the remaining 20\% as  {\small \textsf{Test Requests}} \cite{baeza2015predicting}. 
The thresholds of each prediction algorithm are set to the same values  used in the original techniques. 
Recall that our goal is to show the \emph{feasibility} of small models 
on mobile platforms. 
In turn, this will enable  fine-grained customization of the models to improve accuracy, e.g., tuning the thresholds, pre-processing the training data, and considering additional context. 

Due to the complexity induced by PPM's context-sensitivity (recall Section \ref{sec:foundation}), it was unable to output results when using our entire dataset. To enable a fair comparison among the algorithms, we thus excluded  certain domains that displayed low percentages of repeated requests to reduce the dataset.
We explored multiple cut-off points 
as the repeated percentage at which a domain is excluded 
and found that 10\% was sufficient to enable PPM. 
This eliminated certain users  since all of their requests were excluded, resulting in 9,751 users and 39,004 corresponding models built with the four algorithms.
Interestingly, this process uncovered a potential correlation between the cut-off point's size and the increase in the  models' accuracy. 
This  suggests  a smaller, better-tailored training set can yield even more accurate  models, directly motivating RQ$_3$. 

\subsubsection{\textbf{Accuracy of Prediction Models}}
\begin{figure}[t]
	\centering
	\includegraphics[width=0.4\textwidth]{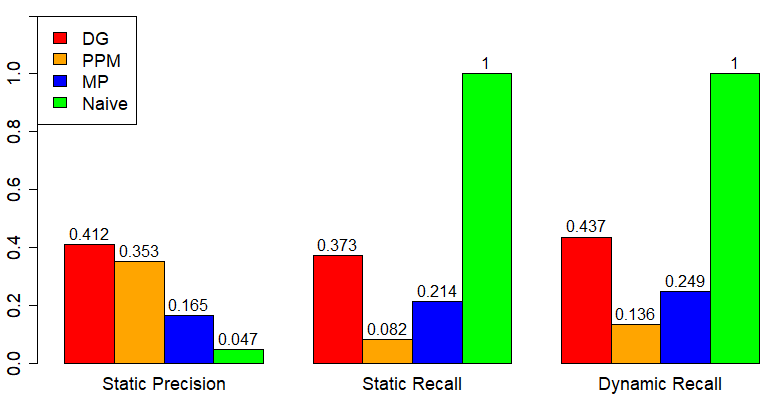}
	\vspace{-4mm}
	\caption{Average values of the 
three accuracy
	metrics across the four algorithms} 
	\vspace{-4mm}
\label{fig:mean_domainFiltered}
\end{figure}

Figure~\ref{fig:mean_domainFiltered} shows the average accuracy results of the 39,004 models.
 Overall, DG outperforms PPM and MP along all three measures, while PPM and MP trade off  precision and recall. 

\looseness-1
As discussed above, Na\"ive is our baseline and it achieves 100\% recall. 
On the other hand, its precision  is poor since it aggressively prefetches every request that has appeared in the past. 
A trade-off between precision and recall should clearly be considered when deciding on a prefetching strategy, based on the specific scenario.
For instance, if the cache  is sufficiently large, Na\"ive can be used to maximize recall.

DG's Static Precision is comparable to the results of traditional large models from the browser domain, where a model is considered to perform well with a precision between 40\% and 50\%~\cite{davison2002predicting,Bouras2004MP,Markatos1996AT1,nanopoulos2001effective}. 
This shows DG's potential on mobile platforms since it achieved comparable precision using much smaller models. 
On the other hand, MP achieved poorer  Static Precision than DG and PPM. 
This was counter-intuitive since MP only uses the most popular requests, which should have yielded smaller model sizes and smaller denominators in the Static Precision formula (recall Section~\ref{sec:study:metric}). 
However, our results suggest that the most popular  requests stop reappearing at some point, 
indicating the need to optimize MP by tuning 
what training data to use and for how long. 


The  Static Recall and Dynamic Recall follow similar trends, meaning that the numbers of \emph{hit} and \emph{miss} requests do not significantly  affect the usefulness of the prediction models.
All three existing algorithms showed  improvements in Dynamic Recall compared to Static Recall, indicating that the {hit} requests they predicted  are usually accessed multiple times.
This directly 
motivates future work on designing effective caching strategies.

\subsubsection{\textbf{Efficiency of Prediction Models}}
\label{rq2:efficiency}

Due to our study's scale, 
we needed to train over 7 million models (detailed in Section~\ref{sec:result:cost}), which required the use of a powerful computing environment. 
 Our experiments were run in parallel on a server with 32 2.60GHz Intel Xeon E5-2650 v2 CPUs and 125GB RAM. 
The average running times of  DG, PPM, MP, and Na\"ive  were 42ms, 431ms, 4ms,  and 9ms \emph{per model},  respectively. 
Note that PPM is at least an order-of-magnitude slower than other algorithms.
This confirmed our earlier observation regarding the scalability issues introduced by PPM's context-sensitivity.

\begin{figure}[t]
	\centering
	\includegraphics[width=0.43\textwidth]{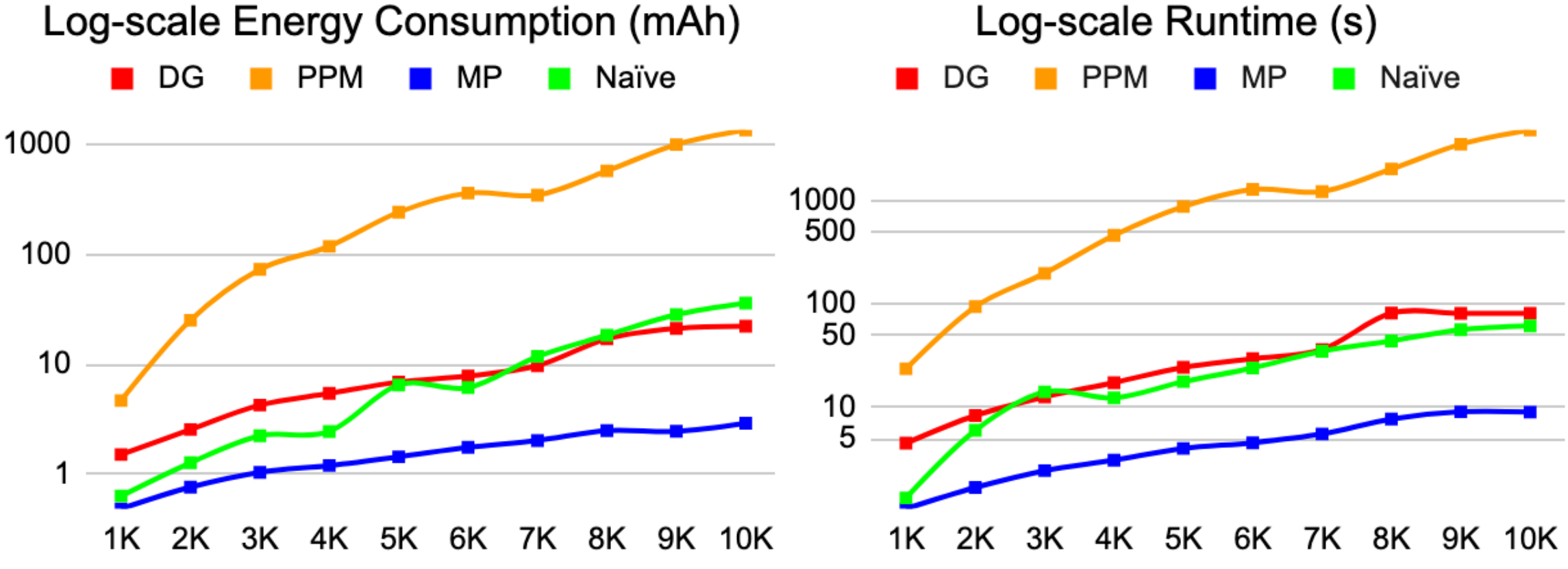}
	\vspace{-4mm}
	\caption{Resource consumption of the four algorithms with ten sets of   different-sized models trained on a mobile device}
\label{fig:energy}
	\vspace{-3.5mm}
\end{figure}

To get further insights into the  models' efficiency, we trained models of 10 different sizes  from $\approx$1,000 to $\approx$10,000 requests, and measured their resource consumption on a mobile device (Honor Play 3 running Android 9.0). 
The sizes were selected empirically, as the resource consumption below 1,000 requests is negligible and training 10,000 requests is already expensive. 
For each size, we selected 10 users whose numbers of requests are the closest to the size and built models with all four algorithms, yielding the results of 400 prediction models.


Figure~\ref{fig:energy} presents the trends (at logarithmic scale) of the models' energy consumptions (in milliAmpere hour) and runtimes (in seconds) as  the number of requests increases. 
The energy consumption is measured by the built-in system tool.
Each data point represents the average resource consumption when training the model of a specific size 10 times 
(the need to train a model multiple times is  discussed in Section~\ref{sec:result:cost}). 
Our results show that DG, MP, and Na\"ive are practical since the largest resource consumption among them is 36mAh~(Na\"ive) and 81s~(DG) when training with $\approx$10,000 requests.\footnote{Modern mobile-devices' battery capacity is $\approx$2,000--5,000mAh~\cite{Michaels2020Apr, Rioja2018Jan}.} MP is by far the most efficient algorithm: it consumes  $<$3mAh and  $<$10s in the worst-case.
By contrast, PPM is not  practical except with the smallest models: 
it begins to surpass the other algorithms' \emph{worst-case} when trained with $\approx$2,000 requests, while its own worst-case consumption is prohibitive at 1,335mAh and 4,740s. 

\looseness-1
Recall from Table \ref{tbl:req_per_user} that  users from our initial dataset average 1,320  requests. 
At first blush, this may suggest that the above analysis was unnecessary and that all four  algorithms should be efficient enough. 
However, given the wide variations across individual behaviors, many users would  likely benefit from larger models. 
Thus, our analysis 
can provide insights for future
work that targets models of varying sizes for specific user groups.

\subsection{RQ3 --   Training-Data Size Reduction}
\label{sec:result:cost}

\begin{itemize}
	\item \textbf{RQ$_3$} -- Can the training-data size be reduced without significantly sacrificing the prediction models' accuracy?
\end{itemize}

To answer RQ$_3$, we first investigate the three data-pruning strategies introduced in Section~\ref{sec:rq:rq3}:  Most Occurring Requests~(MOR), Most Accessed Domains (MAD), and Most Suitable Domains (MSD). 
We then explore  whether there is a \emph{lower-bound}~on the training-data size that yields the smallest prediction models without sacrificing accuracy. 

\subsubsection{\textbf{Data-Pruning Strategies}}
\label{sec:results:pruning}

We apply each of the three pruning strategies to the training data of all 39,004 models studied in RQ$_2$, and use \tool to evaluate the 117,012 pruned  models.
Table~\ref{tbl:pruning} shows the average training-data size reduction  (\emph{Size Red.}) and the average values of our accuracy metrics---Static Precision (\emph{SP}), Static Recall (\emph{SR}), and Dynamic Recall (\emph{DR})---after applying each pruning strategy across all four algorithms;
since the SR and DR values are always 1 for our baseline algorithm Na\"ive, we omit them.
In all but six cases,  accuracy is  improved after data pruning (shaded  cells).

\begin{figure}[b]
	\centering
\includegraphics[width=0.41\textwidth]{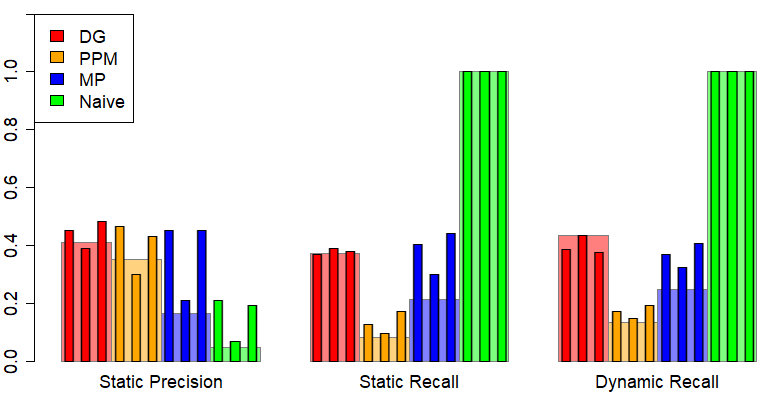}
        \vspace{-3mm}
	\caption{Average accuracy values across the DG, PPM, MP, and  Na\"ive algorithms after applying MOR (left), MAD (middle), and MSD (right), overlayed on top of the original accuracy values from Figure~\ref{fig:mean_domainFiltered}}
\label{fig:pruning}
\end{figure}

Figure~\ref{fig:pruning} overlays the average values from Table~\ref{tbl:pruning} on top of the original results~from Figure~\ref{fig:mean_domainFiltered}.
Each set of three bars corresponds to the results of~the given algorithm after applying  MOR ({left}), MAD ({middle}), and MSD ({right}). For example, the leftmost three bars (in red) represent the MOR-, MAD-, and MSD-yielded values for  DG's Static Precision. 

All three strategies show promising results across all accuracy metrics and  algorithms while reducing the training-data sizes. 
The largest accuracy drop is only 0.06 (DG's Dynamic Recall after applying MSD).
By contrast, accuracy is significantly improved in most cases, with the largest boost of 0.29 (MP's Static Precision after applying MSD).
Note that the largest accuracy drop and boost are both achieved by MSD, suggesting that a pruning strategy may have highly variable impact on different  metrics and/or algorithms.
This is confirmed by the other two strategies: MOR and MAD produce both  improvements and drops for different cases.

\begin{table}[b]
        \vspace{-4mm}
\caption{Average training-data size reduction after applying the MOR, MAD, and MSD pruning strategies, and the resulting Static Precision~(SP), Static Recall (SR), and Dynamic Recall (DR) }
\vspace{-2.5mm}
\label{tbl:pruning}
\centering
\resizebox{\linewidth}{!}{
\begin{tabular}{c|c|c|c|c|c|c|c|c|c|c|c|}
\multicolumn{2}{c|}{}     
                                                                                                                             & \multicolumn{3}{c|}{\textbf{DG}}                                      & \multicolumn{3}{c|}{\textbf{PPM}}                                                             & \multicolumn{3}{c|}{\textbf{MP}}                                                              & \textbf{Na\"ive}                \\ \hline
\multicolumn{1}{|c|}{\textbf{\begin{tabular}[c]{@{}c@{}}Pruning\\ Strategy\end{tabular}}} & \textbf{\begin{tabular}[c]{@{}c@{}}Size\\ Red.\end{tabular}} & SP                            & SR                            & DR    & SP                            & SR                            & DR                            & SP                            & SR                            & DR                            & SP                            \\ \hline
\multicolumn{1}{|c|}{MOR}                                                                 & 54\%                                                         & \cellcolor[HTML]{C0C0C0}0.453 & 0.368                         & 0.386 & \cellcolor[HTML]{C0C0C0}0.468 & \cellcolor[HTML]{C0C0C0}0.128 & \cellcolor[HTML]{C0C0C0}0.173 & \cellcolor[HTML]{C0C0C0}0.452 & \cellcolor[HTML]{C0C0C0}0.404 & \cellcolor[HTML]{C0C0C0}0.368 & \cellcolor[HTML]{C0C0C0}0.212 \\ \hline
\multicolumn{1}{|c|}{MAD}                                                                 & 27\%                                                         & 0.391                         & \cellcolor[HTML]{C0C0C0}0.390 & 0.437 & 0.302                         & \cellcolor[HTML]{C0C0C0}0.097 & \cellcolor[HTML]{C0C0C0}0.150 & \cellcolor[HTML]{C0C0C0}0.212 & \cellcolor[HTML]{C0C0C0}0.302 & \cellcolor[HTML]{C0C0C0}0.325 & \cellcolor[HTML]{C0C0C0}0.069 \\ \hline
\multicolumn{1}{|c|}{MSD}                                                                 & 62\%                                                         & \cellcolor[HTML]{C0C0C0}0.484 & \cellcolor[HTML]{C0C0C0}0.379 & 0.376 & \cellcolor[HTML]{C0C0C0}0.432 & \cellcolor[HTML]{C0C0C0}0.174 & \cellcolor[HTML]{C0C0C0}0.194 & \cellcolor[HTML]{C0C0C0}0.453 & \cellcolor[HTML]{C0C0C0}0.443 & \cellcolor[HTML]{C0C0C0}0.409 & \cellcolor[HTML]{C0C0C0}0.194 \\ \hline
\end{tabular}
}
\vspace{-3mm}
\end{table}

MOR (left) and MSD (right)  outperform MAD (middle) in most cases, while achieving significantly larger size reductions  (Table \ref{tbl:pruning}). 
However, it would be inappropriate to select  the ``best strategy'' \emph{a priori}. 
For instance, if one aims to maximize DG's Dynamic Recall, MAD may in fact be the best pruning strategy. 
If reducing the data size (i.e., resource consumption) is critical, then MSD should be selected, with DG or MP as candidate choices of algorithm. 
Our results provide empirical data to guide  future techniques on selecting suitable pruning strategies and prediction algorithms.

\looseness-1
Among the four algorithms, DG still achieves competitive accuracy, but it benefits the least from data pruning. 
A possible reason is that DG tracks dependencies among requests when building the model and  pruning may cause the loss of dependency information. 
By contrast, MP's accuracy is markedly improved by pruning in all cases. 
MP consistently benefits the most from the MSD strategy that also achieves the largest size reduction (recall Table \ref{tbl:pruning}).
This suggests that MSD's grouping of requests in a domain may have aligned especially well with MP's notion of popular requests.
Finally, it is notable that after the pruning, MP achieves comparable results to DG (the best-performing algorithm identified in RQ$_2$); MP's Static Recall even surpasses DG's after applying either MOR or MSD. 
This observation---that an algorithm may outperform its previously superior competitor if the training set is \emph{reduced}---has not been made previously~\cite{wang2012far,lymberopoulos2012pocketweb,wang2015earlybird,de2007web,ban2007online}. 

\subsubsection{\textbf{Lower-Bound Identification}}
\label{sec:results:sw}

\begin{figure}[b]
        \vspace{-5.75mm}
	\centering
\includegraphics[width=0.4235\textwidth]{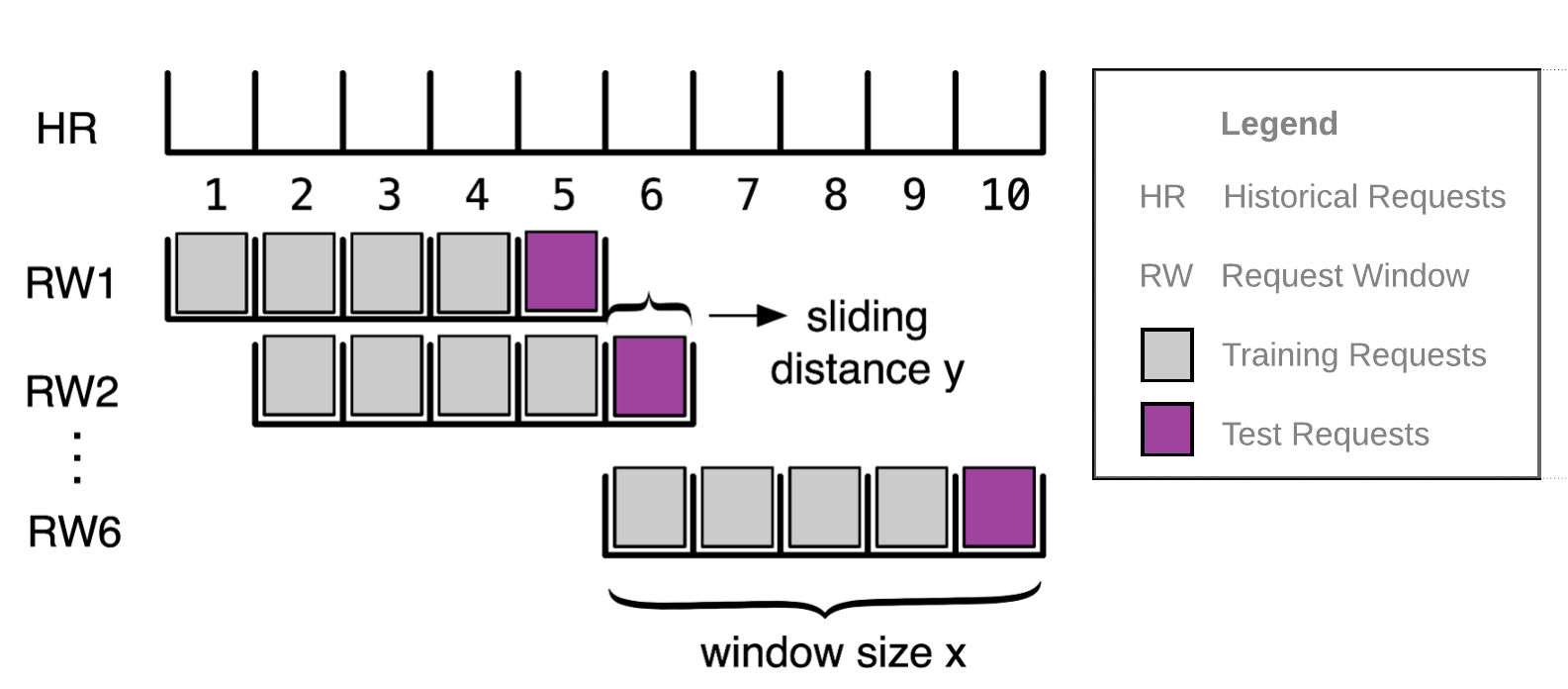}
        \vspace{-4mm}
	\caption{A schematic of the Sliding Window approach with window size $5$, sliding distance $1$, and training ratio $0.8$}
\label{fig:sw}
\end{figure}

Motivated by the above findings, we dove deeper into the relationship between the number of requests in a training set and the accuracy of the resulting prediction models, aiming to explore the \emph{lower-bound} of the training-data size that  yields the smallest model without sacrificing its accuracy.
To do so, we developed a \emph{Sliding Window} (SW) approach for selecting a subset of  training data that, unlike the pruning discussed above, is independent of the data's contents. 
SW tailors \tool's \emph{Training Selection} and \emph{Testing Selection} components. It assesses models of various sizes over different time slices, accounting for various user behaviors throughout the day and ensuring the \emph{freshness} of {\small \textsf{Historical Requests}} used to train the models.
SW allowed us to expand our dataset from 39,004  models studied in RQ$_2$ 
 to over 7 million  models with different training-data sizes of interest. 

Figure \ref{fig:sw} illustrates SW on 10  {\small \textsf{Historical Requests}}. 
We define a \emph{Request Window} (RW) as a subset of {\small \textsf{Historical Requests}} that is used to train and evaluate a prediction model of a certain size. Each RW has a corresponding \emph{window size} that indicates the number of requests in the RW. 
In Figure \ref{fig:sw},  RW has a  window size of 5.
Given {\small \textsf{Historical Requests}} and a window size $x$, we first build a model using the first $x$ requests, and then slide  RW by a \emph{sliding distance} $y$ to build the next  model.
The sliding distance is adjustable. In our study, we set it to be the length of the test set, so that all  {\small \textsf{Test Requests}} from the previous RW can be included in the training set of the next  model. In Figure~\ref{fig:sw}, the sliding distance is 1.
We iteratively build  models of the same window size until the {\small \textsf{Historical Requests}} no longer has sufficient requests to form a complete RW of size $x$. 


We use SW to explore various training ratios and window sizes. 
For brevity, we report the results with the training ratio of  0.8 and 11 window sizes; other values for the two parameters yielded qualitatively similar results. Specifically, we use window sizes of 50, 100, 200, 300, 400, 500, 600, 700, 800, 900, and 1000 requests. This choice of parameters resulted in 1,788,648 prediction models  for individual users with each of the four prediction algorithms, placing the total number of prediction models we studied at over 7.1 million.

\begin{figure}[t]
	\centering
	\includegraphics[width=0.38\textwidth]{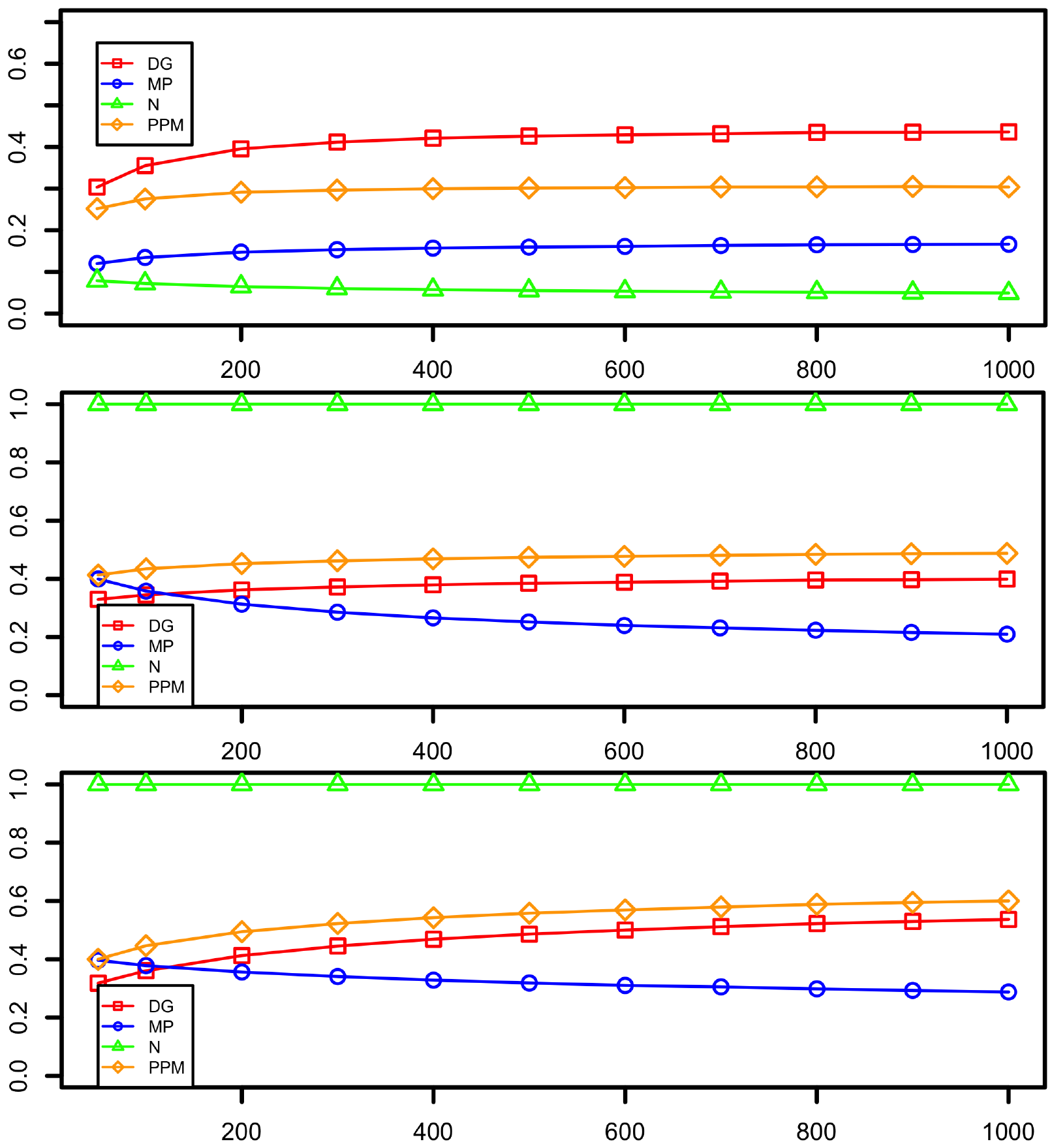}
	\vspace{-3mm}
	\caption{The averages for Static Precision (top), Static Recall~(middle), and Dynamic Recall (bottom).  Y-axes capture the metrics' values; X-axes indicate the different data points corresponding to window sizes}
\label{fig:sw_means}
	\vspace{-4mm}
\end{figure}

\looseness-1
To more closely track the trends of different accuracy metrics as the window sizes increase, we grouped the models of each window size and calculated the mean accuracy values in each group.
Figure~\ref{fig:sw_means} shows the trends of the three metrics across the four algorithms.
Notably, all  trends converge at a certain window size. 
Furthermore, every trend is monotonic, 
 suggesting that there may exist a \emph{cut-off point} after which including more training data will not affect the given model's accuracy. 

To confirm this finding with statistically-significant evidence, 
we conducted pairwise-comparison analysis using the ANOVA post-hoc test based on the Games-Howell method~\cite{gamesAhowell} since it does not require groups of equal sample size. 
This test analyzes whether each pair of groups' means has a statistically-significant difference. 
In our case, 11 window-size groups yield the results of 55 possible pairs for each of the three  metrics calculated based on a given algorithm.
For instance, the 55 pairwise comparison results of DG's Static Precision show that there is a statistically-significant difference among the pairs with window sizes~$\leqslant$~400, but not for sizes~$>$~400. 
Therefore, DG's Static Precision converges at window size 400. 
This is consistent with DG's plot in Figure \ref{fig:sw_means}'s top diagram.

\begin{table}[t]
\centering
\caption{Pairwise comparison result summary}
\vspace{-2mm}
\label{tbl:pairwise}
\small
\centering
\resizebox{0.8\linewidth}{!}{
\begin{tabular}{cccc}
\hline
\textbf{Algorithm} & \textbf{Metric}  & \textbf{Cut-off Point} & \textbf{Trend} \\ \hline
DG                 & Static Precision & 400                    & Positive                  \\ \hline
DG                 & Static Recall    & 500                   & Positive           \\ \hline
DG                 & Dynamic Recall   & 800                    & Positive                  \\ \hline
PPM                & Static Precision & 300                    & Positive                  \\ \hline
PPM                & Static Recall    & 500                & Positive                  \\ \hline
PPM                & Dynamic Recall   & 800                    & Positive                  \\ \hline
MP                 & Static Precision & 400                    & Positive                  \\ \hline
MP                 & Static Recall    & 800                    & Negative                  \\ \hline
MP                 & Dynamic Recall   & 600         & Negative                  \\ \hline
Na\"ive              & Static Precision & 500                    & Negative                  \\ \hline
\end{tabular}
}
\vspace{-4mm}
\end{table}

Table~\ref{tbl:pairwise} summarizes all the pairwise comparisons, including the cut-off points and trends. 
The trends refer to the directionality of the relationship between the means and window sizes: a trend is positive if the mean is higher for larger window sizes, and negative otherwise. 
For cases with positive trends, a cut-off point corresponds to the amount of training data that yielded the highest accuracy; adding more  data beyond this point did not improve the  model's predictive power. In the three cases with negative trends,
 the amount of data  that yielded the highest accuracy corresponds to  the smallest window size (50).

All cut-off points are lower than the largest window size (1000 requests). 
In fact, none of our models needed to be trained with more than 800 requests; on average, the models needed to  be trained by fewer than 400 requests to achieve results comparable to those trained by up to 1000 requests. 
This goes against the conventional wisdom that a prediction model should invariably perform better with more training data, and directly supports our hypothesis stated above.

\section{Broader Implications} 


To our knowledge, our work is the first to provide evidence on the feasibility of history-based prefetching on mobile platforms, using sufficiently small models. 
We have  demonstrated the effectiveness of existing algorithms when properly configured, directly challenging
the previous conclusion that history-based prefetching is ineffective on mobile platforms (avg. 16\% precision and 1\% recall)~\cite{wang2012far}. 
Our study thus motivates re-opening this research area and highlights the opportunity to revisit, and possibly improve, existing prediction algorithms. 
We now discuss the insights gained from our study. 

Even though DG yielded the best accuracy and MP the lowest resource consumption, we argue that neither should be used without a suitable data-pruning strategy since pruning had the dual-benefit of reducing the training-data size and improving the models' accuracy.
We showed that, with an effective pruning strategy, MP can outperform DG, achieving comparable accuracy while maintaining superior resource consumption. 

\looseness-1
Note that the existing algorithms were not built with mobile users in mind, and we applied them \emph{as-is}.
Our results thus can  be treated as the ``floor'' achievable by the algorithms, with a range of possible improvements based on mobile-user characteristics, such as by applying our data-pruning strategies. 
The MSD strategy was the standout overall, with greatest training-data size reduction (avg. 62\%) and accuracy improvement (avg.  84\%).
At the same time, our data showed that certain users benefited more from the MOR or MAD strategies. 
This is consistent with the diverse user behaviors we observed, suggesting future work to categorize  those behaviors and tailor existing or devise new prefetching strategies accordingly. 
Although this paper aims to draw general conclusions based on a large user base, our dataset contains the detailed {\small \textsf{Test Results}} of each individual subject user \cite{repo}, providing a starting point to explore user-behavior categorization. 

\looseness-1
The identified cut-off points for different prediction models (recall Table \ref{tbl:pairwise}) can serve as a guide for exploring suitable model sizes in future techniques. 
For example, to maximize the values of all three accuracy metrics in the two best-performing algorithms---DG and MP---no more than 800 requests are needed.
Recall Figure~\ref{fig:energy}, this indicates that both algorithms will be able to train hundreds~(DG) or even thousands (MP) of models  ``on-device'' with negligible resource consumption. 
In turn, this directly facilitates training multiple models  throughout the day, necessary to maintain the models' ``freshness''. 

DG and MP, the best-performing history-based algorithms, achieved comparable accuracy to the state-of-the-art content-based technique PALOMA (avg.  precision 0.478)~\cite{zhao2018leveraging}.\footnote{Other content-based  techniques did not report  accuracy results \cite{malavolta2019nappa,choi2018appx}.}
Together with the reported limitations of content-based approaches (recall Section \ref{sec:intro}), this highlights a potential advantage of history-based prefetching, as it is applicable to any app. 
However, our data indicates that certain users may not benefit from history-based prefetching since they tend not to send sufficient numbers of repeated requests. 
This suggests an opportunity of combining history-based and content-based approaches to address their respective limitations.
For instance, a content-based technique can analyze the program structure to determine possible subsequent requests when the historical data is limited; on the other hand, a history-based technique can personalize the content-based technique to only prefetch the most likely requests based on an individual user's past behaviors.

Besides the lessons learned from our study, the \tool framework 
 provides a novel, reusable, and tailorable foundation for automatically exploring varioius aspects of history-based prefetching with any dataset of interest. 
Those aspects include identifying suitable training-data sizes, 
trading-off different  metrics, exploring data-pruning strategies, assessing different prediction algorithms, fine-tuning various thresholds, and so on. 
We have already demonstrated how \tool can be used to explore some of these aspects in a flexible manner via ``plugging and playing'' its customizable components. 
In fact, \tool's applicability is not limited to mobile users: it can be applied in other settings by simply replacing its input (i.e., {\small \textsf{Historical Requests}}) with any historical data of interest.
\section{Threats to Validity}
\label{sec:threats}


Our dataset was collected from a university's network with 11,476 \textbf{users}, suggesting that our findings may not hold for all settings and types of users. 
However, universities have large and diverse populations,  spanning students, faculty,  administrative staff, contract employees, outside vendors, and visitors.  
The university also provides on-campus housing, dining, working, and entertainment venues, which captured user behaviors in different environments. 
Our user base is also over 400 times larger than the closest comparable study~\cite{wang2012far}.

Our data was collected over a \textbf{single day}, raising the possibility that our findings may not apply to historical data collected over  longer periods.
However, this was done by design: we aim to investigate the feasibility of \emph{small prediction models} on mobile platforms, to mitigate the challenge of obtaining large amounts of user data as discussed in Section \ref{sec:intro}.
\cut{ Moreover, as mentioned above, the sampled population tended to access the network at different times of day, in different ways, and with different intensities.} 

The collected data includes \textbf{network traffic} from both mobile apps and mobile browsers across different device types. Our results may thus fail to capture specific characteristics of mobile apps, mobile browsers, or certain devices. However, our goal is to demonstrate the {feasibility} of history-based prefetching 
	on mobile platforms in general.
As discussed in Section \ref{sec:result:cost}, a smaller but better-tailored  model may yield better results, thus our study ``underapproximates'' the model's achievable accuracy.
While we may have missed certain users who never connected to the university network, we had access to data from a large number of diverse users, which allowed us to obtain statistically-significant evidence for our results.

We assumed that the \textbf{cache} size  is unbounded and {cached requests} do not expire.
This is guided by our objective to assess the \emph{accuracy} of \emph{small} prediction models 
(recall Section~\ref{foundation:focus}).
\section{Conclusion and Future Work}
\label{sec:conclusion}

\looseness-1
This paper presents the first attempt to investigate the feasibility of history-based prefetching using small models on mobile platforms. 
We did so by developing \tool, a tailorable framework that enabled us to automatically assess over 7 million models built from real mobile-users' data. 
Our results provide empirical evidence of the feasibility of small prediction models, opening up a new avenue for improving mobile-app performance while
meeting stringent privacy requirements. 
We further show that existing algorithms from the browser domain can produce reasonably accurate and efficient models on mobile platforms, and provide several insights on how to improve them.
For example, we developed several strategies for reducing the training-data size while maintaining, even increasing, a model's accuracy.
Finally, \tool's reusability and customization provide a flexible foundation for subsequent studies to further explore various aspects of history-based prefetching. 

While this initial study focused on identifying general trends 
that span our large user base, tracking personalized network-usage patterns across different time periods (e.g., morning vs. night, weekday vs. weekend, work vs. vacation) is likely to result in even more accurate  models. 
This may require access to significantly more user data, however---in volume, variety, and geographic span---than we have currently been granted. We will work on overcoming this challenge, and on the related critical issue of user privacy inherent in studies such as ours. 

\section*{Acknowledgment}
This work is supported by the U.S. National Science Foundation under grants 1717963, 1823354, and 2030859 (the Computing Research Association for the CIFellows Project), U.S. Office of Naval Research under grant N00014-17-1-2896, and the National Natural Science Foundation of China under grants 62072046 and 61702045.

\clearpage
\bibliographystyle{IEEEtran}
\bibliography{LessIsMore.bib}

\end{document}